\documentclass{article}

\usepackage{PRIMEarxiv}

\usepackage[utf8]{inputenc} 
\usepackage[T1]{fontenc}    
\usepackage{hyperref}       
\usepackage{url}            
\usepackage{booktabs}       
\usepackage{amsfonts}       
\usepackage{nicefrac}       
\usepackage{microtype}      
\usepackage{lipsum}
\usepackage{fancyhdr}       
\usepackage{graphicx}       
\graphicspath{{media/}}     
\usepackage[version=4]{mhchem}
\DeclareUnicodeCharacter{2212}{-}
\begin{document}
\noindent\begin{minipage}{17cm}
\rule{\textwidth}{0.7mm}
    \begin{center}
{\huge\bfseries Strain driven conducting domain walls in a Mott insulator \par}
\vspace{0.1cm}
    \end{center}
\rule{\textwidth}{0.7mm}\par
\vspace*{1ex}
        \end{minipage}
        
\vspace{4mm}
L. Puntigam$^1$,  M. Altthaler$^1$, S. Ghara$^1$, L. Prodan$^1$, V. Tsurkan$^{1,2}$, S. Krohns$^1$, I. Kézsmárki$^1$, D. M. Evans$^1$\\

$^1$ Experimental Physics V, Center for Electronic Correlations and Magnetism, Institute of Physics, University of Augsburg, 86135 Augsburg, Germany \\
$^2$ Institute of Applied Physics, MD 2028 Chisinau, Moldova\\
Email: donald.evans@uni-a.de\\
\vspace{4mm}

\begin{abstract}
Rewritable nanoelectronics offers new perspectives and potential to both fundamental research and technological applications. Such interest has driven the research focus into conducting domain walls: pseudo-2D conducting channels that can be created, positioned, and deleted \textit{in-situ}. However, the study of conductive domain walls is largely limited to wide-gap ferroelectrics, where the conductivity typically arises from changes in charge carrier density, due to screening charge accumulation at polar discontinuities. This work shows that, in narrow-gap correlated insulators with strong charge-lattice coupling, local strain gradients can drive enhanced conductivity at the domain walls – removing polar-discontinuities as a criteria for conductivity. By combining different scanning probe microscopy techniques, we demonstrate that the domain wall conductivity in \ce{GaV4S8} does not follow the established screening charge model but rather arises from the large surface reconstruction across the Jahn-Teller transition and the associated strain gradients across the domain walls. This mechanism can turn any structural, or even magnetic, domain wall conducting, if the electronic structure of the host is susceptible to local strain gradients - drastically expanding the range of materials and phenomena that may be applicable to domain wall-based nanoelectronics.
\end{abstract}


\section{Introduction}
Extreme miniaturisation of active electronic components, while keeping the complexity of their function,\cite{moore-1965} has spurred much interest in low-dimensional objects, including emergent and artificially designed 2D interfaces\cite{Hwang2012, Catalan-2012} and flakes of 2D materials.\cite{Geim2007, Sujay2016} For instance, naturally occurring structurally sub-nanometer wide interfaces, e.g. domain walls, have been reported to posses the same inherent electronic response as existing circuit elements, such as switches\cite{mundy-2017} and half-wave rectifiers.\cite{schaab-2018} In addition, ferroelectric domain walls can be reconfigured \textit{in-situ} by a variety of external fields which can lead to exotic bulk responses. Such bulk responses offer the opportunity to both enhance existing technology (e.g. magnetoresistance,\cite{ghara-2021} colossal dielectric constants,\cite{puntigam-2021} memristive behaviour\cite{McConville-2020}) but also provide next generation functionalities like, negative capacitance,\cite{guy-2021} above band gap photovoltaic effects,\cite{bhatnagar-2013} and domain wall nanoelectronics.\cite{meier-2021} Such effects have been discussed from both a fundamental and a technological perspective in recent reviews.\cite{evans-2020, salje-2021}\\

For all of these bulk responses, the key requirement is for the domain walls to exhibit a different conductivity compared to the surrounding material. Therefore, much of the research has focused on ferroelectrics as the build up of screening charges at domain walls with polar discontinuities are known to modify the local conductivity.\cite{meier-2021,evans-2020} Examples of this, in single crystals, includes \ce{BaTiO3},\cite{Sluka-2013} \ce{(Ca,Sr)3Ti2O7},\cite{Oh-2015}  Cu-Cl boracite’s,\cite{McQuaid-2017} \ce{LiNbO3},\cite{Schroder-2012} h-\ce{$R$MnO3} ($R$ representing rare earth metals),\cite{Meier-2012} and \ce{GaV4S8}.\cite{ghara-2021} Because of the energetically costly nature of such polar discontinuities, their spontaneous formations are normally restricted to improper ferroelectrics.\cite{Bednyakov-2018, Evans-2020c} Indeed, so established is this screening charge mechanism that it is surprising for an improper ferroelectric material, exhibiting polar discontinuities, not to have conducting domain walls.\cite{Bak-2020} In this scenario, the type of domain wall that is expected to have enhanced conductivity depends on the electronic structure of the host material: In a p-type (n-type) semiconductor the tail-to-tail (head-to-head) domain walls attract screening holes (electrons) and thus provide enhanced conductivity relative to the bulk.\cite{Holstad-2018, Jiang-2018, Rojac-2017} The corresponding head-to-head (tail-to-tail) wall in a p-type (n-type) material is then expected to have reduced conductivity compared to the bulk.\cite{meier-2021, Bednyakov-2018, Evans-2020c, holstad-2020}\\

Note, in ferroelectrics, particularly thin-films, further conductivity mechanisms have been reported.\cite{mundy-2017, Eliseev-2012, Seidel-2009, Morozovska-2012, Maksymovych-2012, Stolichnov-2015} But it is challenging, especially in oxides, to disentangle intrinsic effects and those associated with, e.g., enhanced defect density at domain walls,\cite{Guyonnet-2011, Farokhipoor-2011} which can change surface Schottky barriers and hence conductivity.\cite{schaab-2018} There have also been reports of domain wall conductivity and even superconductivity in non-ferroelectrics. Examples of the former case include, the antiferromagnetic insulators with conducting magnetic domain walls - attributed to the presence of mid-gap states,\cite{Ma-2015} or in VO$_{2}$ domain walls around the phase transition, as the new phase nucleates preferentially at ferroelastic domain walls (i.e. a metal state on heating, and an insulating state on cooling).\cite{Tselev-2010} Superconductivity has reported to appear at structural twin walls before the bulk transition,\cite{Aird-1998, Evans-2019} attributed to the local symmetry of the walls and the corresponding strain coupling.\\

On the bulk scale it is long established that the conductivity of highly correlated systems, such as Mott insulators, can be change by changing the inter-atomic distances. Practically, such changes to the inter-atomic distance can be achieved via applied strain, hydrostatic-, or chemical- pressures.\cite{Imada-1998, Guzman-Verri-2019}
In this work, we move these bulk concepts to the nanoscale by showing how highly conducting domain walls can be created using local strain gradients, rather than traditional screening charge approach. Taking the narrow band gap Mott insulator \ce{GaV4S8} as a representative of such highly correlated systems, we use scanning probe microscopy to reveal that the enhanced conductivity arises at domain walls which have a large surface reconstruction, a signature of local strain gradient. This mechanism explains, among other things, the origin of indistinguishable enhanced conductivities at “head-to-head” and “tail-to-tail” domain walls, which cannot be explained via the conventional screening-charge-based mechanism. Our results provide a new mechanism of generating conductivity at domain walls in single crystals, expanding the field of domain wall nanoelectronics to material systems, like Mott insulators, whose band structures are strongly influenced by strain gradients.

\section{The lacunar spinel \ce{GaV4S8}}
For this proof-of-principle study \ce{GaV4S8} is an ideal template system: it is characterized by a correlation band gap of $\sim$0.3\,eV  and has a strong charge-lattice coupling.\cite{Reschke-2020, Geirhos-2021, Hu-2021} The material crystallizes in a face centered cubic (NaCl-like) structure comprising [\ce{V4S4}] heterocubane entities and [\ce{GaS4}] tetrahedra, as depicted in \textbf{Figure\,\ref{fig:1}\,a}.\cite{Pocha-2000} In lacunar spinels, the strong charge-lattice coupling arises from the Jahn-Teller instability of the heterocubane cluster,\cite{Pocha-2000, Johrendt-1998} and in \ce{GaV4S8} this drives the ferroelectric transition.\cite{Pocha-2000, Hlinka-2016, Butykai-2017} \\

At the Jahn-Teller transition of \ce{GaV4S8}, taking place at \ce{T_{JT}}$\approx$42\,K, the symmetry is reduced from cubic ($F\bar{4}3m$) to polar rhombohedral ($R3m$) via an elongation of the \ce{V4S4} units along one of the cubic $\langle$111$\rangle$ directions.\cite{Pocha-2000} For the four possible domains, \ce{\textbf{P}1}-\ce{\textbf{P}4}, depicted in Figure\,\ref{fig:1}\,b, this distortion leads to a ferroelectric polarization of $\text{P}_{\text{s}}$ = 2.3\,\textmu C/\ce{cm^2}.\cite{ghara-2021, Ruff-2015} Due to the lack of $\pm P$ domain pairs, $180^\circ$ domain walls of neighboring domains cannot emerge in \ce{GaV4S8}, where the largest polar discontinuities normally occur.\cite{McConville-2020, Mosberg-2019} The four polarisation directions are illustrated schematically in Figure\,\ref{fig:1}\,b, and the origin of the ferroelectricity is discussed in detail in Refs.~\cite{Geirhos-2021,ghara-2021} and the references therein. Below \ce{T_{JT}}, the material undergoes a magnetic ordering at $\sim$13\,K, with several competing magnetic phases, including a N\'eel-type skyrmion lattice phase.\cite{Kezsmarki-2015} It is also well documented that the electronic structure of \ce{GaV4S8} is very sensitive to inter-atomic spacing, and can be modified with either hydrostatic pressure,\cite{Mokdad-2019, Wang-2021} or chemical pressure.\cite{Kim-2014} \\

\section{Results and Discussion}
\begin{figure}[hth]
  \centering
  \includegraphics[width=0.6\linewidth]{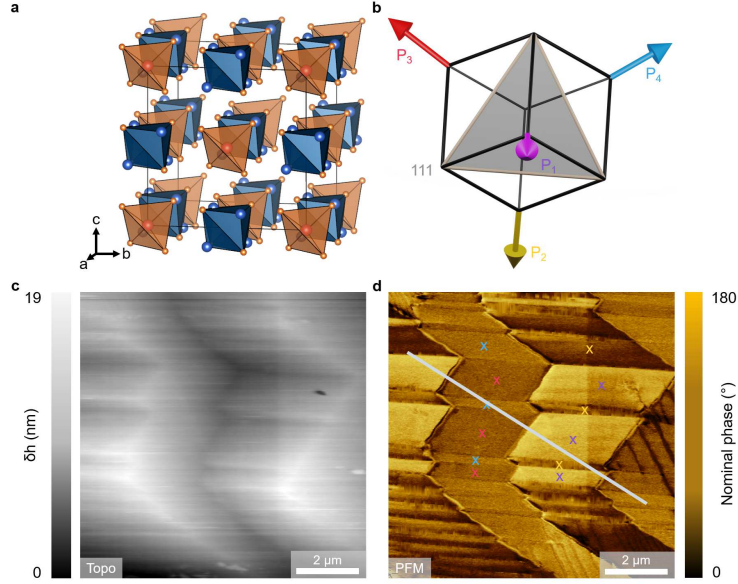}
  \caption{\textbf{a} Schematic unit cell of \ce{GaV4S8} (lacunar spinel, chemical structure [\ce{AM4X8}]) consisting of [\ce{M4X4}] heterocubane entities (blue) and [\ce{AX4}] tetrahedra (red) with the space group $F\bar{4}3m$. \textbf{b} Schematic showing the four different polarization directions with respect to the cubic crystal system. The gray triangle represents the (111) plane which was the measured surface in this work. \textbf{c}  Topography information of the as-grown (111) surface of a \ce{GaV4S8} single crystal taken at 30\,K. Bright colors show increased height, while dark colors are lower areas. The lamella structure of the domains, as well as the domain wall position, is visible in the topography. \textbf{d} Out-of-plane phase information from a PFM scan of the same area as \textbf{c}, the golden areas show domains with pure \ce{\textbf{P}1} polarization direction, while the brown and mixed areas show \ce{\textbf{P}2}-\ce{\textbf{P}4} polarized domains. The orientation of the polarization vectors is indicated by the colored cross, where the direction of the colors is defined in \textbf{b}. The scans were performed using a boron doped single crystalline diamond tip (BdSC).}
   \label{fig:1}
\end{figure}

To understand and correlate the local structural and electronic properties at domain walls in \ce{GaV4S8}, we use scanning probe microscopy performed on an as-grown (111) surface of a single crystal. Figure\,\ref{fig:1}\,c displays a representative topography image recorded at 30\,K. The topography shows a large amount of surface reconstruction below the Jahn-Teller transition, which is manifested as a series of ridges and valleys that zig-zag across the surface. A direct comparison of the topography above and below the transition is given in Figure\,S1, and some of the implications from such surface reconstruction are described in Supporting Note 1.\\

The corresponding out-of-plane piezo response force microscopy (PFM) phase image allows us to identify these regions as different ferroelectric domains, Figure\,\ref{fig:1}\,d. Since the image is taken on a (111) surface, the polarization of one of the four domain states, \ce{\textbf{P}1}, is perpendicular to this surface, while the polarization for the other three span 71$^{\circ}$ to the scanned surface. Hence, \ce{\textbf{P}1} is assigned to the domain with largest amplitude in the out-of-plane piezo response, i.e. the brightest areas in Figure\,S2. Domains with lighter shades of brown correspond to domain states \ce{\textbf{P}2} - \ce{\textbf{P}4}, which can be distinguished by considering the orientation of the mechanically compatible domain walls,\cite{Sapriel-1975} and described for \ce{GaV4S8} in Ref. \cite{ghara-2021}. This is summarised by the different colored crosses in Figure\,\ref{fig:1}\,d, which follow the color convention of Figure\,\ref{fig:1}\,b. The large topographic features, the ridges and valleys, clearly correlate with the ferroelectric domain structure resolved by PFM, with the larges changes in topography arising at $\{$110$\}$-type domain walls. From symmetry arguments these $\{$110$\}$-type domain walls are expected to have polar-discontinuities and consequentially positive and negative bound charges. We will referrer to these walls as ``p-type" or``n-type" when information about the charge state is important, and otherwise as ``zig-zag" domain walls.\\

\begin{figure}[hth]
\center
  \includegraphics[width=0.45\linewidth]{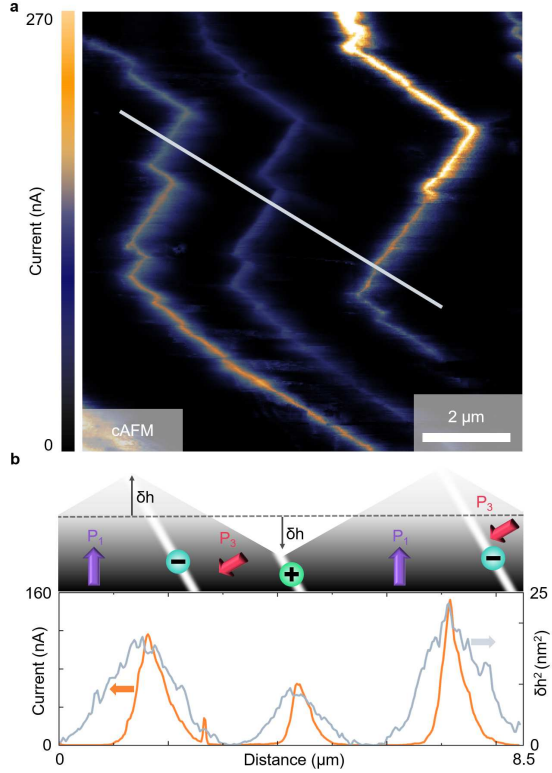}
  \caption{\textbf{a} cAFM image showing conducting zig-zag-domain walls in \ce{GaV4S8}, where bright colors show increased conductivity. \textbf{b} Cross-sections of the cAFM data, taken along the light gray line in \textbf{a} shown with the change in current across three domain walls. A schematic addition of the associated topography (gray background), the polarization direction (arrows) and the inferred screening charge type along the domain walls (bright lines with positive and negative signs) is also added. In addition, the measured current is correlated to the observed change in topography across the Jahn-Teller transition, see Supplementary Note 1. Topography and PFM data are derived from Figure\,\ref{fig:1}\,. Note: the direction of the schematic domain walls is only illustrative. The scans was performed using a BdSC diamond tip.}
  \label{fig:2}
\end{figure}

In order to see which of the domain walls are associated with increased conductivity, we collect conductive atomic force microscopy data (cAFM) (with 60\,V), \textbf{Figure\,\ref{fig:2}\,a}, of the region illustrated in Figure\,\ref{fig:1}. In the cAFM image the bright gold colors are areas of highly enhanced current (values up to 270\,nA) and the dark areas are regions of lower current. These areas of enhanced current align with the zig-zag-domain wall pattern observed in the topography and PFM images of Figure\,\ref{fig:1}\,c and d, respectively. The observed current values, of several hundred nanoampers, are high for domain walls in ferroelectrics, which typically have values in the pico- to nanoampere range.\cite{meier-2021} One of the most striking features of the cAFM data in Figure\,\ref{fig:2} is that both the head-to-head and the tail-to-tail domain walls, which are expected to posses opposite bound charge, exhibit enhanced conductivity. From the conventional polar discontinuity argument, this would mean that ``n-type" and ``p-type" charge carriers have the same effects on the conductivity; in contrast to the literature on bulk \ce{GaV4S8} \cite{Janod-2015}. This is also surprising for semiconducting ferroelectrics in general, as typically the walls with an accumulation of majority carriers is expected to have enhanced conductivity, while the walls with an accumulation of minority carriers have a reduced conductivity relative to the bulk.\cite{Catalan-2012, Evans-2020c, holstad-2020, Nataf-2020} In addition, if adjacent domain walls had majority p-type and n-type charge carriers, the intersection of the wall would form a depletion region. The absence of any such region in Figure\,\ref{fig:2}\,a is inconsistent with the polar discontinuity model but consistent with the strain-gradient driven model.\\

To provide a more qualitative comparison between conduction and topographic, we plot cross-sections of both in Figure\,\ref{fig:2}\,b, indicated by the grey line in Figure\,\ref{fig:2}\,a, and schematically add the topography and ferroelectric polarization vectors. We observe that large current values arise at both head-to-head and tail-to-tail domain walls, accompanied by a surface reconstruction ($\delta$h$^2$) - discussed below and detailed in Supporting Note 1. The cAFM cross-section shows that the conductive regions are relatively wide, extending ca. 750 nm away from the center of the wall. This non-local nature is consistent with a strain- gradient around the wall, which effects the surrounding electronic structure of the material. However, this is inconsistent with the idea of polar discontinuity driven domain wall conductivity, as this is necessarily limited to the wall itself. For reference, in materials where the polar discontinuity driven mechanisms is well established, such as the hexagonal manganites,\cite{Evans-2020c, Holtz-2017} Cu-Cl boracite,\cite{McQuaid-2017} \ce{BaTiO3},\cite{Crassous-2015} or \ce{BiFeO3},\cite{Farokhipoor-2011} the walls typically have an effective electronic width of ca. $\sim$100\,nm.\\

\begin{figure}[hth]
\center
  \includegraphics[width=0.6\linewidth]{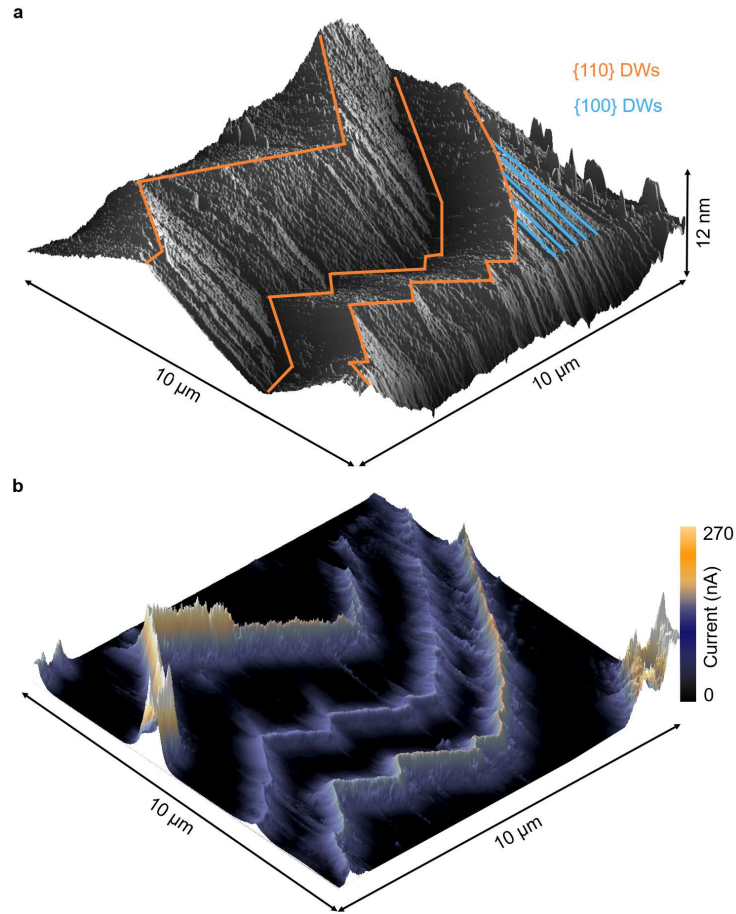}
  \caption{\textbf{a} 3D plot of the topography information of a series of zig-zag-domain walls. The z-axis is the height, the orange lines depict the position of $\{$110$\}$-type domain walls and representative blue lines the $\{$100$\}$ domain walls. \textbf{b} 3D plot of the cAFM response of the area from \textbf{a}. The large hills and valleys in topography, that follow the $\{$110$\}$-type domain walls, are associated with enhanced conductivity (bright colors) suggesting a common origin. The scans were performed using a BdSC diamond tip.}
  \label{fig:3}
\end{figure}

In order to understand the correlation of topography and conductivity in \ce{GaV4S8} further, we consider the large surface reconstruction that arises during the Jahn-Teller transition. To represent these local strain-gradients, using an approach inspired by Aizu's work on spontaneous strains across a symmetry lowering structural transition,\cite{Aizu-1970} we compare the changes in topography relative to an assumed surface. This represents the crystal surface had the transition not taken place. Practically, this is achieved by taking the square of the difference in height ($\delta$h$^2$) of the topography versus a smooth plane, schematically shown by the dashed grey line in Figure\,\ref{fig:2}\,b. This change in height, taken along the cross-section of Figure\,\ref{fig:2}\,a, gives an excellent correlation with the measured current (Figure\,\ref{fig:2}\,b).\\

In order to visualize the topography/conductance correlation is ubiquitous over the observed area, we present 3D maps of the topography and conductivity in \textbf{Figure\,\ref{fig:3}}. These are the same data sets provided in Figure\,\ref{fig:1}\,c and Figure\,\ref{fig:2}\,a. This 3D representation shows that the large peaks and valleys in the topography along the $\{$110$\}$-type domain walls (orange lines) are associated with areas of enhanced conductivity (bright colors), while small changes in surface reconstruction at the $\{$100$\}$ domain walls (such as those highlighted with blue lines) exhibit no observed change in conductivity.\\

\begin{figure}[hth]
\center
  \includegraphics[width=0.7\linewidth]{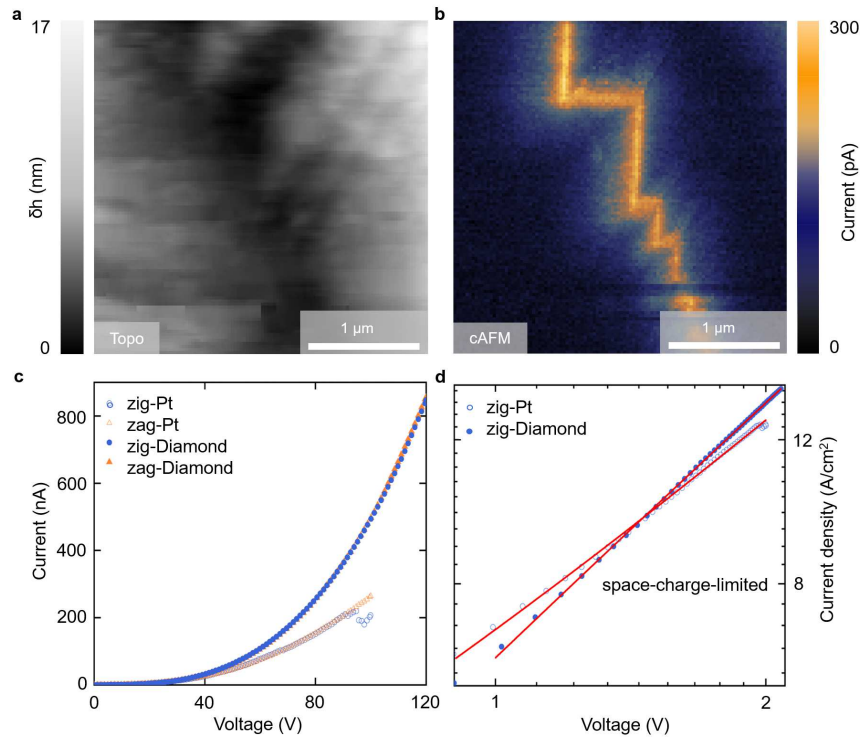}
  \caption{\textbf{a} Topography information of the as-grown (111) surface of a \ce{GaV4S8} single crystal taken at 30\,K with Pt-Si tip. Bright colors show increased height, while dark colors are lower areas. The lamella structure of the domains, as well as the domain wall position, is visible next to the valley of the zig-zag-domain walls in the topography. \textbf{b} cAFM image of the region in \textbf{a}, bright colors indicate areas with enhanced conductivity. \textbf{c} Current response to an applied voltage at the zig- and zag-domain walls, taken with both Pt-Si and BdSC diamond tips. \textbf{d} Double logarithmic plot of current density versus applied voltage. Red lines are fits with a space-charge-limited model giving rise to a high quality factor, $R^2$.}
  \label{fig:4}
\end{figure}

To further test the idea that the conductivity is enhanced by local strain gradient induced changes in band structure, rather than the accumulation of screening charges, we consider the system in terms of classical semiconductor-metal interface (a Schottky barrier). In a semiconductor-metal interface the conductivity depends on the respective work functions, in this case the work function of the cAFM tip (e.g. Pt-Si and BdSC-diamond) and that of the sample. Natural, areas with p-type conductivity would have a different band structure to areas of n-type conductivity. Therefore, we  performed cAFM and PFM measurements using a metallic Pt-Si tip (\textbf{Figure\,\ref{fig:4}}) in addition to the measurements with a BdSC diamond tip (Figures\,\ref{fig:1} - \ref{fig:3}). The topography of the same sample is given in Figure\,\ref{fig:4}\,a, and the corresponding cAFM data is presented in Figure\,\ref{fig:4}\,b. The topography shows the same structure of zig-zag-domain pattern observed in Figure\,\ref{fig:1}. Two things are apparent: the observed current values are smaller, hundreds of picoamps compared to hundreds of nanoamps (c.f. Figure\,\ref{fig:2}\,a), and both zig- and zag- walls have similar conductivity. This suggests that, while the tip work function changes the values of the Schottky barrier (and therefore the measured current), there is no difference in band structure between the zig- and zag- domain walls, as this would change the relative current across the domain walls. While highly problematic for the polar discontinuity driven mechanism, it is consistent with strain gradient induced changes to the band structure, as this would have carriers of the same type at both walls.\\

In order to identify the conductivity mechanism, we collect I(V)-spectroscopy data across zig- and zag- walls. The plotted curves in Figure\,\ref{fig:4}\,c, for both types of tip and both types of domain wall, are the average values collected over ten data sets. The current response for each tip is the same for both type of domain walls. We fit these voltage dependent responses with the established models of conductivity in semiconductors: Schottky emission, Fowler-Nordheim tunneling, direct tunneling, thermionic-field emission, Poole-Frenkel emission, hopping conduction and space-charge-limited conduction. Following the methodology of Shih 2014,\cite{Shih-2014} the I(V) data is plotted in a way that, if the model describes the data well, the plot will be linear. Such a linear behaviour is only observed for the space-charge-limited representation of the I(V) data, Figure\,\ref{fig:4}\,d. This does not change with different tip material. The other plots, corresponding to the different models, are given in supplementary Figure\,S3.\\

\begin{figure}[hth]
\center
  \includegraphics[width=0.9\linewidth]{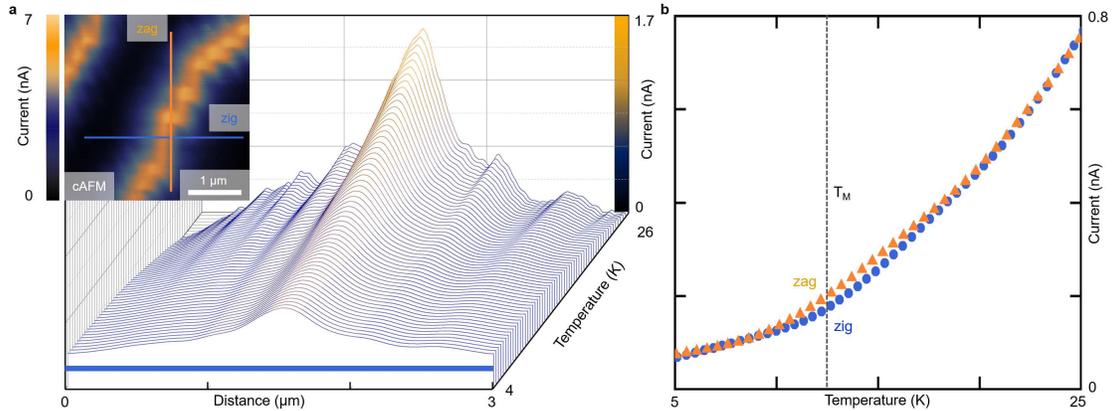}
  \caption{\textbf{a} Waterfall plot of the current response from 26\,K down to 4\,K of a zig-domain wall. The inset shows a cAFM measurement to indicate the location of the line scans for zig- (blue) and zag- (orange) domain walls, collected at 30\,K. \textbf{b} Temperature-dependent change in peak current values of both zig- and zag-domain walls, extracted from the waterfall plots represented in \textbf{a}. The scans were done using a BdSC diamond tip.}
  \label{fig:5}
\end{figure}

To exclude the possibility that the equal conductance of zig- and zag-domain wall segments, as already observed using two types of tips, is an accident caused by a special combination of carrier densities and mobilities, we investigate the temperature dependence of the domain wall conductivity. It is established, that even in the simplest semiconducting materials, such as Si - where the electrons can be described as non-interacting particles - electrons and holes have different temperature dependent mobilities; in the case of Si, \ce{T^{-2.6}} and \ce{T^{-2.3}}, respectively.\cite{Morin-1954} In Mott insulators, the electrons are highly correlated and cannot be considered non-interacting. If such classical 'electron' and 'hole' type charge carriers could form, a difference in their respective mobilities is therefore expected. Separate cAFM line scans across a zig- and a zag- wall (see inset) were collected while cooling from 26\,K to 4\,K, and the data for the zig-wall is given by the waterfall plot of \textbf{Figure\,\ref{fig:5}\,a}. We plot the peak values of currents from our temperature dependency for both walls in Figure\,\ref{fig:5}\,b. The temperature dependent conductivity is indistinguishable for the two walls, except around the magnetic ordering transition, suggesting the same charge carrier type at both walls. An anomaly around the magnetic ordering is expected, if the walls have different magnetic properties, a discussion of which is the subject of forthcoming work.\\

\section{Conclusion}
 We use AFM topography images to show that a large surface reconstruction arises during the Jahn-Teller transition, which coincides with the formation of ferroelectric domain walls, identified with PFM. The corresponding cAFM images show that both the head-to-head and tail-to-tail domain walls are highly conducting compared to the bulk, and this conductivity is directly proportional to the amount of surface reconstruction squared.  Such surface reconstruction is naturally associated with strain gradients, which in Mott insulators, or any other system where the electronic structure is sensitive to strain, will change the band structure and therefore the conductivity. We confirm the strain gradient model, as the temperature-dependent current response of the zig- and zag-domain walls is indistinguishable, even for cAFM tips of different work function. Furthermore, I(V)-spectroscopy identifies that the conductivity is bulk limited and well explained by space charge limited conductivity. This mechanism could turn any structural or even magnetic domain walls conducting, if the electronic structure of the host is susceptible to strain, and so significantly broaden the scope of domain wall physics by expanding the range of materials that could be used for domain wall nanoelectronics.\\
 
 \section{Experiment}
\ce{GaV4S8} single crystals have been grown by the chemical transport reactions method. As starting material for growth  preliminary synthesized polycrystalline powder was used, the polycrystals were prepared by solid state reactions from the high-purity elements: Ga (99.9999\%) (Alfa Aesar, Ward Hill, USA), V (99.5\%) (Alfa Aesar, Ward Hill, USA) and S (99.999\%) (Strem Chemicals Inc., Newburyport, USA). The iodine was utilized as the transport agent in the single crystal growth. The growth was performed at temperatures between $800^\circ\text{C}$ and $850^\circ\text{C}$. Perfect truncated octahedron-like samples with dimension up to 5 mm were obtained.\\

The atomic force microscopy (AFM) scans were conducted on the same single crystalline sample at various temperatures using an attoAFM I (attocube systems AG, Haar, Germany) atomic force microscope with Pt-Si-tips (PtSi-FM-10, NanoandMore GmbH, Wetzlar, Germany) as well as boron doped single crystal diamond (AD-2.8-AS, Bruker France, Wissembourg, France) tips. Local transport data was gained by cAFM with varying dc-voltages, ranging from 1\,V to 60\,V applied to the back electrode. Piezoresponse force microscopy (PFM) was conducted with the off-resonant method using frequencies ranging from 19\,kHz up to 99\,kHz with 10\,V excitation voltage.\\

I(V)-spectroscopy was conducted by placing a grid of measurement points over an area of interest. For each point a voltage range from 0\,V to 120\,V (WMA-200 High voltage amplifier, Falco Systems, Katwijk aan Zee, Netherlands) was applied and the respective current signal tracked. The acquired three dimensional data cube is evaluated with an in-house written MatLab-script (The MathWorks, Inc., New Mexico, USA). The phase and amplitude signal of the PFM scans was measured with a lock-in amplifier SR860 (Stanford research systems, Sunnyvale, USA).\\

\medskip
\textbf{Author Contributions}\par
L. Puntigam recorded the scanning probe microscopy data supervised by M. Altthaler and D. M. Evans. V. Tsurkan and L. Prodan have prepared the crystals. L. Puntigam, M. Altthaler, D. M. Evans, S. Krohns and I. Kézsmárki interpreted the data and drafted the manuscript. All authors discussed the results and contributed to the final version of the manuscript.

\section*{Acknowledgments}
L. Puntigam, M. Altthaler, S. Ghara, L. Prodan, V. Tsurkan, S. Krohns, and I. K\'ezsm\'arki acknowledge by the Deutsche Forschungsgemeinschaft (DFG) Grant No. KE 2370/3-1 and the Transregional Research Collaboration TRR 80 (Augsburg, Munich and Stuttgart; Project No. 107745057).\\
M. Altthaler acknowledges funding of the DFG Priority Program SPP2137, Skyrmionics, under Grant No. KE 2370/1-1.\\
L. Prodan and V. Tsurkan also acknowledge the support provided by the project ANCD 20.80009.5007.19 (Moldova).\\ 
D. M. Evans acknowledges the support provided by DFG individual fellowship number (EV 305/1-1).\\

\bibliographystyle{unsrt}  
\bibliography{references}  

\end{document}


\noindent\begin{minipage}{17cm}
\rule{\textwidth}{0.7mm}
    \begin{center}
{\huge\bfseries Strain driven conducting domain walls in a Mott insulator \par}
\vspace{0.1cm}
    \end{center}
\rule{\textwidth}{0.7mm}\par
\vspace*{1ex}
        \end{minipage}
        
\vspace{4mm}
L. Puntigam$^1$,  M. Altthaler$^1$, S. Ghara$^1$, L. Prodan$^1$, V. Tsurkan$^{1,2}$, S. Krohns$^1$, I. Kézsmárki$^1$, D. M. Evans$^1$\\

$^1$ Experimental Physics V, Center for Electronic Correlations and Magnetism, Institute of Physics, University of Augsburg, 86135 Augsburg, Germany \\
$^2$ Institute of Applied Physics, MD 2028 Chisinau, Moldova\\
Email: donald.evans@uni-a.de\\
\vspace{4mm}

\section{Supporting Figures}
\renewcommand{\thefigure}{S\arabic{figure}}
\setcounter{figure}{0} 
AFM topography images denoting the strong surface reconstruction that arises during the Jahn-Teller transition, is shown in \textbf{Figure\,\ref{fig:S1}}. Figure\,\ref{fig:S1}\,a provides the as-grown (111) surface at 60\,K and Figure\,\ref{fig:S1}\,b the same area below \ce{T_{JT}} at 30K. The 30\,K image shows that cooling through the transition leads to a series of ridges and valleys, running approximately perpendicular to the scratches, with a variable periodicity.  

\begin{figure}[htb]
  \centering
  \includegraphics[width=0.35\linewidth]{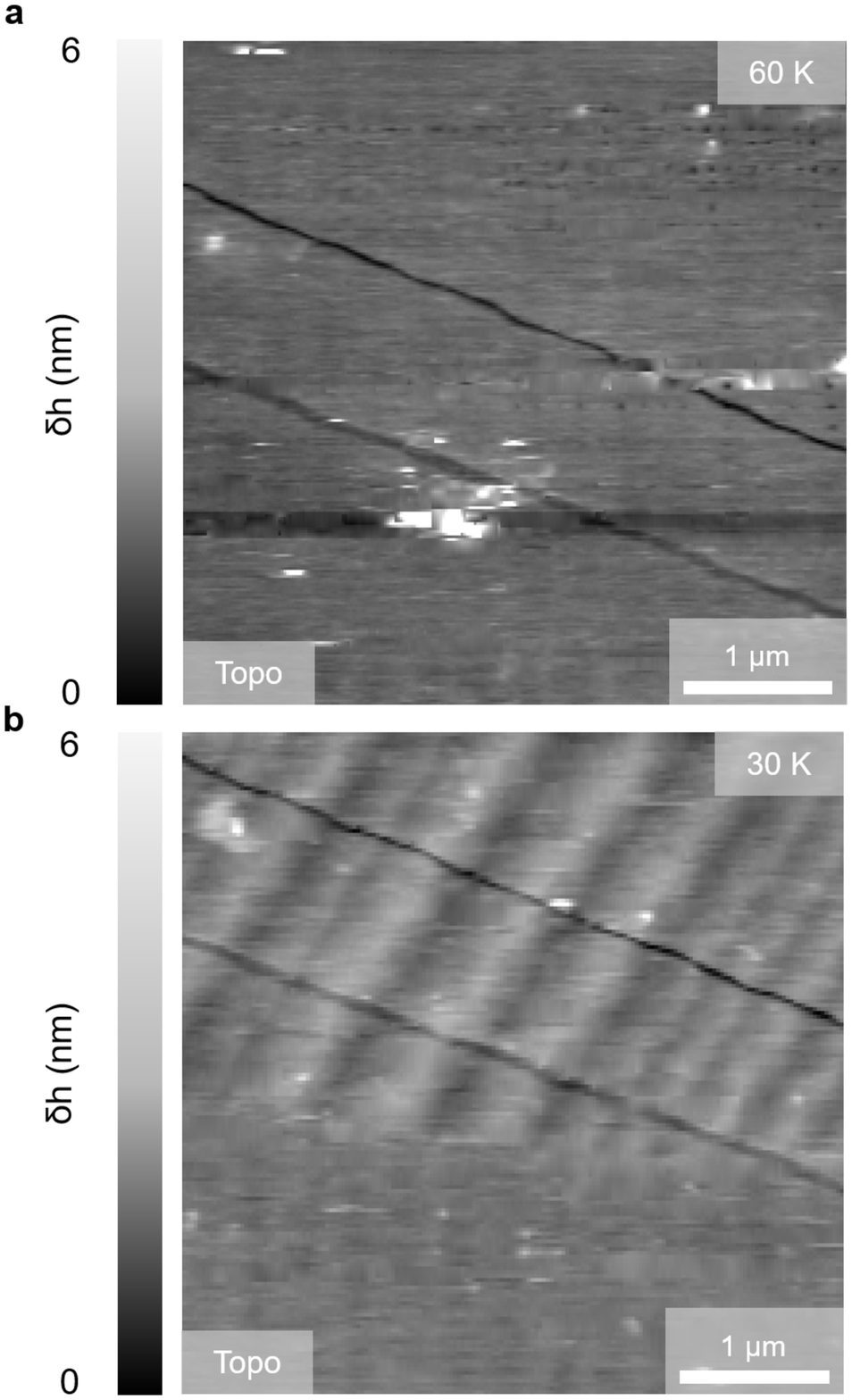}
  \caption{\textbf{a} Topography information of the as-grown (111) surface taken at 60\,K. Bright colors show higher areas, while darker colors lower areas. \textbf{b} The same area scanned at 30\,K, showing a corrugation of the topography into a series of ridges and valleys of ca. 4 nm in height. These scans were performed across two nano-cracks to confirm it is the same area in both images. These scans were collected using a BdSC diamond tip.}
  \label{fig:S1}
\end{figure}
\newpage

\begin{figure}[htb]
  \centering
  \includegraphics[width=0.5\linewidth]{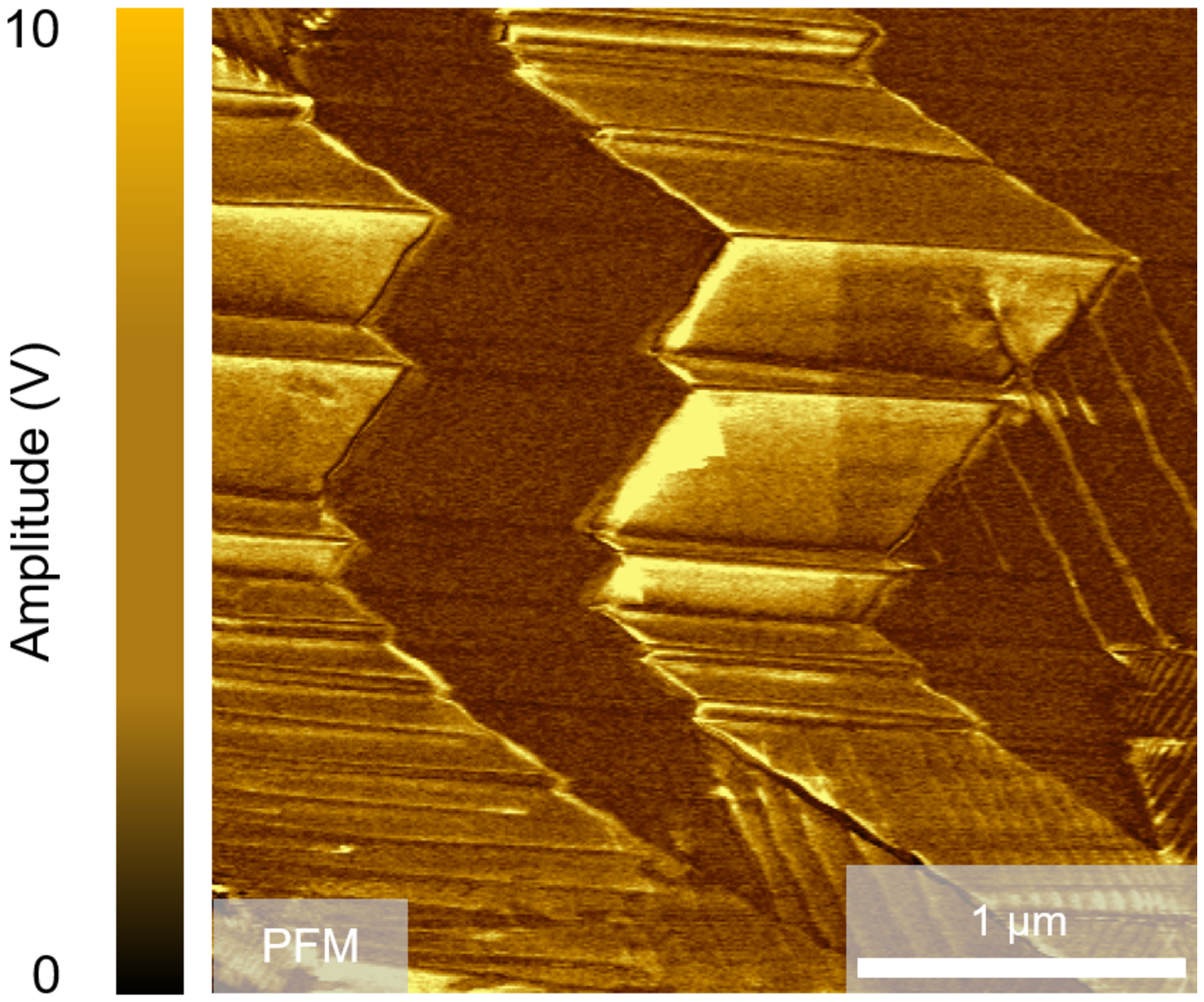}
  \caption{\textbf{a} PFM image showing the respective amplitude information to the scan in Figure\,2\,d. The scans were performed using a BdSC diamond tip.}
  \label{fig:S2}
\end{figure}

\textbf{Figure\,\ref{fig:S2}} is the PFM amplitude to the phase image of Figure\,2\,d of an area with different domains below \ce{T_{JT}}. The largest piezo response is used to label \ce{\textbf{P}1}, the out of plane polarization according to the (111) surface.

\begin{figure}[hth]
  \centering
  \includegraphics[width=0.5\linewidth]{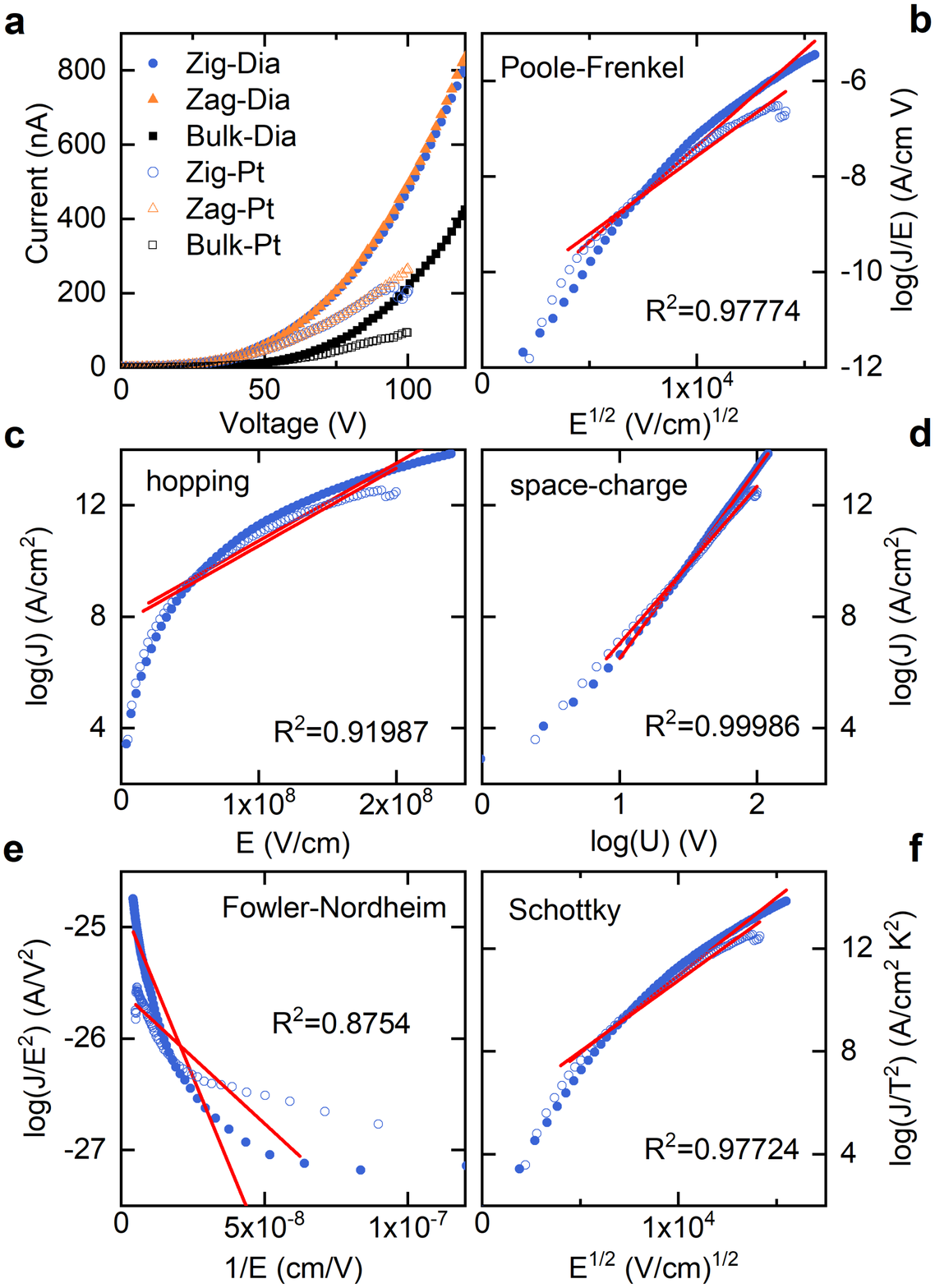}
  \caption{\textbf{a} IV-curves taken at zig- and zag-domain walls and on the domain, with both boron-doped single crystal diamond tips and platinum tips. \textbf{b} Current flow per E-field is plotted over the square root of the E-field in order to fit the Poole-Frenkel model with a linear fit (red line) \cite{Shih-2014}. The data is not linearized meaning it is not a good fit. \textbf{c} Logarithmic plot of the current density over the E-field to linearize the data to a Hopping-model, the resulting plot is not linear and so it is not a good fit. The space-charge model (\textbf{d}) is fitted by using a double logarithmic plot of the current flow over the applied voltage, with $R^2$=0.9986. For this fit the data is linearized meaning it is a good fit. Neither the Fowler-Nordheim model \textbf{e} which is linearized by plotting the logarithm of the current flow over the squared E-field against the inverse E-field, nor the Schottky model (\textbf{f}), plotted with the logarithm of the current flow over the root of the E-field, give a linear fit. This means the domains and domain walls are well described by the space-charge limited conductivity model. Note: the models from \textbf{b}, \textbf{c} and \textbf{d} are bulk-limited conductivity mechanisms, while the models used for \textbf{e} and \textbf{f} are electrode contact limited models.}
  \label{fig:S3}
\end{figure}

\newpage
\textbf{Figure\,\ref{fig:S3}} provides the models, Poole-Frenkel, hopping, space-charge, Fowler-Nordheim and Schottky, fitted to IV spectroscopy data (Figure\,\ref{fig:S3}\,a). A linearization indicates a perfect match, which is achieved for space-charge limited process depicted in the main manuscript (Figure\,4\,d). For comparison reasons the same data is plotted in Figure\,\ref{fig:S3}\,d.

\newpage
\section{Supplementary Notes}
Supplementary note 1.\\ 
Aizu's work on spontaneous strain correlates the strain to the difference between the lattice parameters below a phase transition, and the extrapolated values of the lattice parameters from the high temperature phase – i.e. the expected lattice parameters if the transition had not occurred.\cite{Tagantsev-2010, Aizu-1970} Such spontaneous strain occurs in ferroelastic and coelastic materials.\cite{Salje-1991, Wadhawan-2000} We use an analogues approach but, rather than considering the difference between the lattice parameters, we take into account the difference between the surface reconstruction of the low temperature phase relative to the assumed surface of the high temperature phase. Any surface reconstruction is energetically costly for the material and it typical forms along with domain walls to compensate the change in unit cell volume across a phase transition in order to minimise the absolute change in volume of the crystal.\cite{Tilley-2013} Detailed information about strain gradients across structural domain walls can be found in the seminal references \cite{Salje-1991, Wadhawan-2000}, or the pioneering work on ferroelastics and twin boundaries (in terms of an elastic continuum theory) by Barsch and Krumhansl.\cite{Barsch-1984} \\

The above concepts allow to calculate a value for $\delta$h$^2$ which, via elastic continuum theory, is used as a proxy for the local strain gradients. Note: we plot the squared value, rather than any other even power, as it is the lowest order term that describes the symmetry.\\